\documentclass[twocolumn,prd,showpacs,showkeys,showkeywords,preprintnumbers,amsmath,amssymb]{revtex4}

\usepackage{graphicx}
\usepackage{dcolumn}
\usepackage{bm}
\usepackage[english]{babel}

\newcommand{\DM}   {dark matter}

\newcommand{\msun}{M_{\odot}}

\begin{document}
\title{Search for Gamma-rays from Dark Matter annihilations \\around Intermediate Mass Black Holes with the H.E.S.S. experiment}
\author{F. Aharonian$^{1,14}$}
 \author{A.G.~Akhperjanian$^{2}$}
 \author{U.~Barres de Almeida$^{8}$}
 \altaffiliation[]{supported by CAPES Foundation, Ministry of Education of
 Brazil}
 \author{A.R.~Bazer-Bachi$^{3}$}
 \author{B.~Behera$^{14}$}
 \author{M.~Beilicke$^{4}$}
 \author{W.~Benbow$^{1}$}
 \author{K.~Bernl\"ohr$^{1,5}$}
 \author{G.~Bertone$^{6}$}
  \altaffiliation[]{not a H.E.S.S. member}
\email{bertone@iap.fr}
 \author{C.~Boisson$^{7}$}
% ...

 \affiliation{$^{1}$ Max-Planck-Institut f\"ur
Kernphysik, P.O. Box 103980, D 69029 Heidelberg, Germany}
\affiliation{$^{2}$
 Yerevan Physics Institute, 2 Alikhanian Brothers St., 375036 Yerevan,
Armenia} \affiliation{$^{3}$ Centre d'Etude Spatiale des
Rayonnements, CNRS/UPS, 9 av. du Colonel Roche, BP 4346, F-31029
Toulouse Cedex 4, France} \affiliation{$^{4}$ Universit\"at
Hamburg, Institut f\"ur Experimentalphysik, Luruper Chaussee 149,
D 22761 Hamburg, Germany} \affiliation{$^{5}$ Institut f\"ur
Physik, Humboldt-Universit\"at zu Berlin, Newtonstr. 15, D 12489
Berlin, Germany} \affiliation{$^{6}$ Institut d'Astrophysique de
Paris, UMR 7095 CNRS, Universit\'e Pierre et Marie Curie, 98 bis
boulevard Arago, 75014 Paris, France} \affiliation{$^{7}$ LUTH,
Observatoire de Paris, CNRS, Universit\'e Paris Diderot, 5 Place
Jules Janssen, 92190 Meudon, France} \affiliation{$^{8}$
IRFU/DSM/CEA, CE Saclay, F-91191 Gif-sur-Yvette, Cedex, France}
\affiliation{$^{9}$ University of Durham, Department of Physics,
South Road, Durham DH1 3LE, U.K.} \affiliation{$^{10}$ Unit for
Space Physics, North-West University, Potchefstroom 2520,
    South Africa}
\affiliation{$^{11}$ Laboratoire Leprince-Ringuet, Ecole
Polytechnique, CNRS/IN2P3,
 F-91128 Palaiseau, France}
\affiliation{$^{12}$ Laboratoire d'Annecy-le-Vieux de Physique des
Particules, CNRS/IN2P3, 9 Chemin de Bellevue - BP 110 F-74941
Annecy-le-Vieux Cedex, France} \affiliation{$^{13}$ Astroparticule
et Cosmologie (APC), CNRS, Universit\'e Paris 7 Denis Diderot, 10,
rue Alice Domon et Leonie Duquet, F-75205 Paris Cedex 13, France}
 \altaffiliation[Also at ]{UMR 7164 (CNRS, Universit\'e Paris VII, CEA, Observatoire de
 Paris)}
\affiliation{$^{14}$ Dublin Institute for Advanced Studies, 5
Merrion Square, Dublin 2, Ireland} \affiliation{$^{15}$
Landessternwarte, Universit\"at Heidelberg, K\"onigstuhl, D 69117
Heidelberg, Germany} \affiliation{$^{16}$ Laboratoire de Physique
Th\'eorique et Astroparticules, CNRS/IN2P3, Universit\'e
Montpellier II, CC 70, Place Eug\`ene Bataillon, F-34095
Montpellier Cedex 5, France} \affiliation{$^{17}$ Universit\"at
Erlangen-N\"urnberg, Physikalisches Institut, Erwin-Rommel-Str. 1,
D 91058 Erlangen, Germany} \affiliation{$^{18}$ Laboratoire
d'Astrophysique de Grenoble, INSU/CNRS, Universit\'e Joseph
Fourier, BP 53, F-38041 Grenoble Cedex 9, France}
\affiliation{$^{19}$ Institut f\"ur Astronomie und Astrophysik,
Universit\"at T\"ubingen, Sand 1, D 72076 T\"ubingen, Germany}
\affiliation{$^{20}$ LPNHE, Universit\'e Pierre et Marie Curie
Paris 6, Universit\'e Denis Diderot Paris 7, CNRS/IN2P3, 4 Place
Jussieu, F-75252, Paris Cedex 5, France} \affiliation{$^{21}$
Institute of Particle and Nuclear Physics, Charles University, V
Holesovickach 2, 180 00 Prague 8, Czech Republic}
\affiliation{$^{22}$ Institut f\"ur Theoretische Physik, Lehrstuhl
IV: Weltraum und Astrophysik, Ruhr-Universit\"at Bochum, D 44780
Bochum, Germany} \affiliation{$^{23}$ University of Namibia,
Private Bag 13301, Windhoek, Namibia} \affiliation{$^{24}$
Obserwatorium Astronomiczne, Uniwersytet Jagiello\'nski, ul. Orla
171, 30-244 Krak{\'o}w, Poland} \affiliation{$^{25}$ Nicolaus
Copernicus Astronomical Center, ul. Bartycka 18, 00-716 Warsaw,
Poland} \affiliation{$^{26}$School of Physics \& Astronomy,
University of Leeds, Leeds LS2 9JT, UK} \affiliation{$^{27}$
School of Chemistry \& Physics,
 University of Adelaide, Adelaide 5005, Australia}
\affiliation{$^{28}$ Toru{\'n} Centre for Astronomy, Nicolaus
Copernicus University, ul. Gagarina 11, 87-100 Toru{\'n}, Poland}
\affiliation{$^{29}$ Instytut Fizyki J\c{a}drowej PAN, ul.
Radzikowskiego 152, 31-342 Krak{\'o}w, Poland}
\affiliation{$^{30}$ European Associated Laboratory for Gamma-Ray
Astronomy, jointly supported by CNRS and MPG}
%
%
\date{\today}
\begin{abstract}
The H.E.S.S. array of Cherenkov telescopes has performed, from
2004 to 2007, a survey of the inner Galactic plane at photon
energies above 100 GeV. About 400 hours of data have been
accumulated in the region between -30 and +60 degrees in Galactic
longitude, and between -3 and +3 degrees in Galactic latitude.
Assuming that \DM\, is composed of Weakly Interacting Massive
Particles, we calculate here the H.E.S.S. sensitivity map for
\DM\, annihilations, and
derive the first experimental constraints on the 
$^{\backprime\backprime}$mini-spikes$^{\prime\prime}$ scenario, in
which a gamma-ray signal arises from \DM\, annihilation around
Intermediate Mass Black Holes. The data exclude the proposed
scenario at a 90\% confidence level for \DM\, particles with
velocity-weighted annihilation cross section $\sigma v$ above $\rm
10^{-28}\,cm^3s^{-1}$ and mass between 800 GeV and 10 TeV.
\end{abstract}

\maketitle

\section{Introduction}
A substantial body of cosmological and astrophysical measurements
suggests that $\sim$22\% of the Universe is composed of
non-baryonic \DM\, (DM), e.g.~\cite{spergel}, commonly assumed to
be in the form of Weakly Interacting Massive Particles (WIMPs)
arising in extensions of the Standard Model of Particle Physics
(for recent reviews see~\cite{bergstrom0,bertone0}). The lightest
neutralino arising in supersymmetric (SUSY) extensions of the
Standard Model~\cite{jungman}, and the first excitation of the
Kaluza-Klein bosons, $\tilde{B}\rm ^{(1)}$, in universal extra
dimension (UED) theories~\cite{appelquist,cheng,servant} are among
the most widely discussed \DM\, candidates.

Although accelerator and direct \DM\, searches may well provide
useful hints on the nature of \DM, a correct identification is
likely to require the combination of different techniques,
including the indirect searches, based on the detection of the
annihilation products of \DM\, particles. The annihilation rate
being proportional to the square of the \DM\, density integrated
along the line of sight, regions with enhanced \DM\, density are
primary targets of indirect searches. Among them are the Galactic
halo~\cite{silk}, external galaxies~\cite{baltz0}, galaxy
clusters~\cite{fornengo},
substructures~\cite{calcaneo,tasitsiomi,stoehr,khousiappas0,baltz1,pieri0,koushiappas,diemand,Pieri:2007ir},
and the Galactic Center
(GC)~\cite{aharonian0,zaharijas,mambrini,profumo,cesarini}.

The GC, in particular, has attracted  significant interest. The
distribution of \DM\, at the GC is actually highly uncertain, due
to lack of resolution in N-body simulations, and to the many
astrophysical effects that further complicate the situation, such
as the presence of the supermassive black hole coincident with Sgr
A*, gravitational scattering of \DM\, off the stellar cusp, and
\DM\, annihilation~\cite{bertone1}.

Dwarf spheroidal galaxies in the Local Group~\cite{mateo} have
also been considered  as targets for gamma-ray detection since
they seem to represent dark matter dominated
regions~\cite{evans,profumo1,sanchez,bergstrom,strigari,aharonian1}.
More recently, the Sloan Digital Sky Survey (SDSS) revealed the
existence of new satellites~\cite{willman,irwin,walsh} offering
appealing features for \DM\, searches. Radial velocity dispersion
of stars in Galaxy satellites usually  implies large
mass-to-luminosity ratio. Nevertheless, the lack of accurate
measurements on the velocity dispersion may induce large
systematic effects on the parameters used for the modelling of the
\DM\, halo. Even in case of accurate kinematic data for some of
the Galaxy satellites, only faint annihilation signals are
expected for smooth dark matter halos due to their distance of
$\sim$100 kpc.

Mini-spikes around Intermediate Mass Black Holes (IMBHs) have been
recently proposed as promising targets for indirect \DM\,
detection~\cite{zhao}. Accurate predictions, in the context of
well-defined astrophysical scenarios, have been derived for the
distribution and luminosity of these objects~\cite{bertone34}.
Mini-spikes might in fact be detected as bright point-like sources
in gamma-rays~\cite{bertone3} and neutrinos~\cite{bertone4}, and
the prospects for detection with satellites with large field of
view such as the upcoming GLAST experiment~\cite{glast} to be
launched in 2008 appear particularly promising.

Current Imaging Atmospheric Cherenkov Telescopes (IACTs) can also
effectively search for these objects. The H.E.S.S. (High Energy
Stereoscopic System) experiment has already surveyed a significant
part of the Galactic plane. With the combination of a large field
of view, very good angular resolution and off-axis performance,
H.E.S.S. has reached the sensitivity to accurately map the
Galactic plane in scan-based observations.

In this paper, H.E.S.S. data are used to derive for the first time
experimental exclusion limits on the \DM\, annihilation signals
within the context of the mini-spike scenario. The paper is
organized as follows. Section II is devoted to the mini-spike
scenario and the gamma-ray flux expected from \DM\, annihilations
in mini-spikes. In section III, we present the H.E.S.S. data from
the Galactic plane survey and compute the H.E.S.S. flux
sensitivity map to \DM\, annihilations in the Galactic plane
region. Exclusion limits are then derived for \DM\, annihilation
from mini-spikes. Finally, section IV is devoted to the discussion
of the results obtained in this study.

\section{\label{sec:imbh}\DM\, annihilations in mini-spikes}

\subsection{Dark matter candidates in MSSM and Kaluza-Klein models}

In this paper we focus on two particle physics scenarios beyond
the Standard Model which provide well-motivated WIMP \DM\,
candidates with masses and couplings at the electroweak scale to
account for the non-baryonic \DM. The annihilation of WIMP pairs
can produce in the final state a continuum of gamma-rays whose
flux extends up to the DM particle mass, from the hadronization
and decay of the cascading annihilation products.

\begin{figure}[!ht]
\includegraphics[scale=0.44]{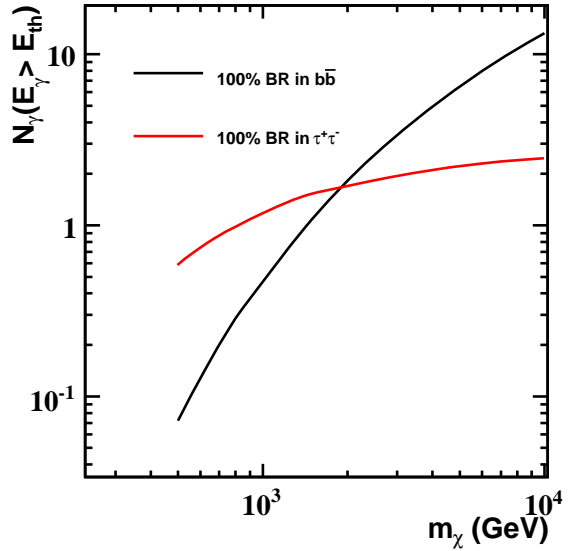}
\caption{Number of gammas per annihilation above the energy
threshold E$_{\rm th}$=~100~GeV, $N_{\gamma}(E_{\gamma}>E_{\rm
th})$, as a function of the neutralino mass, m$_{\chi}$, for 100\%
BR in the $b\bar{b}$ channel (solid black line) and 100\% BR in
the $\tau^+\tau^-$ (solid red line). This number corresponds to
the annihilation spectrum integrated from E$_{\rm th}$ up to
m$_{\chi}$. Below 2 TeV neutralino masses, more gammas per
annihilation are expected in the case of the $\tau^+\tau^-$
channel than for $b\bar{b}$, thus the latter will provide a
conservative value. Above 2 TeV, the $b\bar{b}$ channel yields
more gammas than the $\tau^+\tau^-$ one.}
\label{fig:Integrated_Annihilation_Spectra}
\end{figure}

Minimal Supersymmetric extensions of the Standard Model (MSSM)
predict the lightest supersymmetric particle (LSP) to be stable in
the case of R-parity conserving scenarios~\cite{jungman}. In
various SUSY breaking schemes, this fermionic particle is the
lightest neutralino $\chi$. In general MSSM, the gamma-ray
spectrum from neutralino annihilation is not uniquely determined
and the branching ratios (BRs) of the open annihilation channels
are not determined since the DM particle field content is not
known. Soft and hard spectra for the neutralino pair annihilation
are therefore considered. The annihilation spectra are
parametrized using PYTHIA~\cite{pythia} simulations.
Fig.~\ref{fig:Integrated_Annihilation_Spectra} shows the number of
gammas above the energy threshold $E_{\rm th}$=~100~GeV,
$N_{\gamma}(E_{\gamma}>E_{\rm th})$, as a function of the
neutralino mass $m_{\chi}$. $N_{\gamma}(E_{\gamma}>E_{\rm th})$ is
lower in the case of a 100\% BR $b\bar{b}$ channel than in the
$\tau^+\tau^-$ channel for $m_{\chi}$ below 2 TeV. Above 2 TeV,
more gammas are expected in the $b\bar{b}$ channel.

In some specific scenarios, the branching ratios of the
annihilation channels can be computed given that the field content
of the DM particle is known. In the SUSY scenario dubbed
AMSB~\cite{chamseddine,hall,otha,barbieri} (Anomaly Mediated
Supersymmetry Breaking) arising in SUSY
theories~\cite{randall,guidice}, the DM candidate is the lightest
neutralino with a predominant wino component which annihilates in
$W^+W^-$ pairs  decaying in quarks and leptons. Large annihilation
cross sections are expected which may lead to detectable gamma-ray
fluxes~\cite{hooper}. Within this scenario, neutralino masses may
extend up to tens of TeV. The other \DM\, candidate considered
here arises in theories with universal extra dimensions (UED). In
Kaluza-Klein (KK) scenarios with KK-parity conservation, the
lightest KK particle (LKP) is stable. Most often, the LKP is the
first KK mode of the hypercharge gauge boson, referred hereafter
to as $\tilde{B}^{(1)}$~\cite{appelquist,cheng,servant}. In the
Kaluza-Klein case, the BRs are extracted from~\cite{servant}.
$\tilde{B}^{(1)}$ pairs annihilate mainly into fermion pairs: 35\%
in quark pairs and 59\% in charged lepton pairs.

\begin{figure*}[!ht]
\includegraphics[scale=0.44]{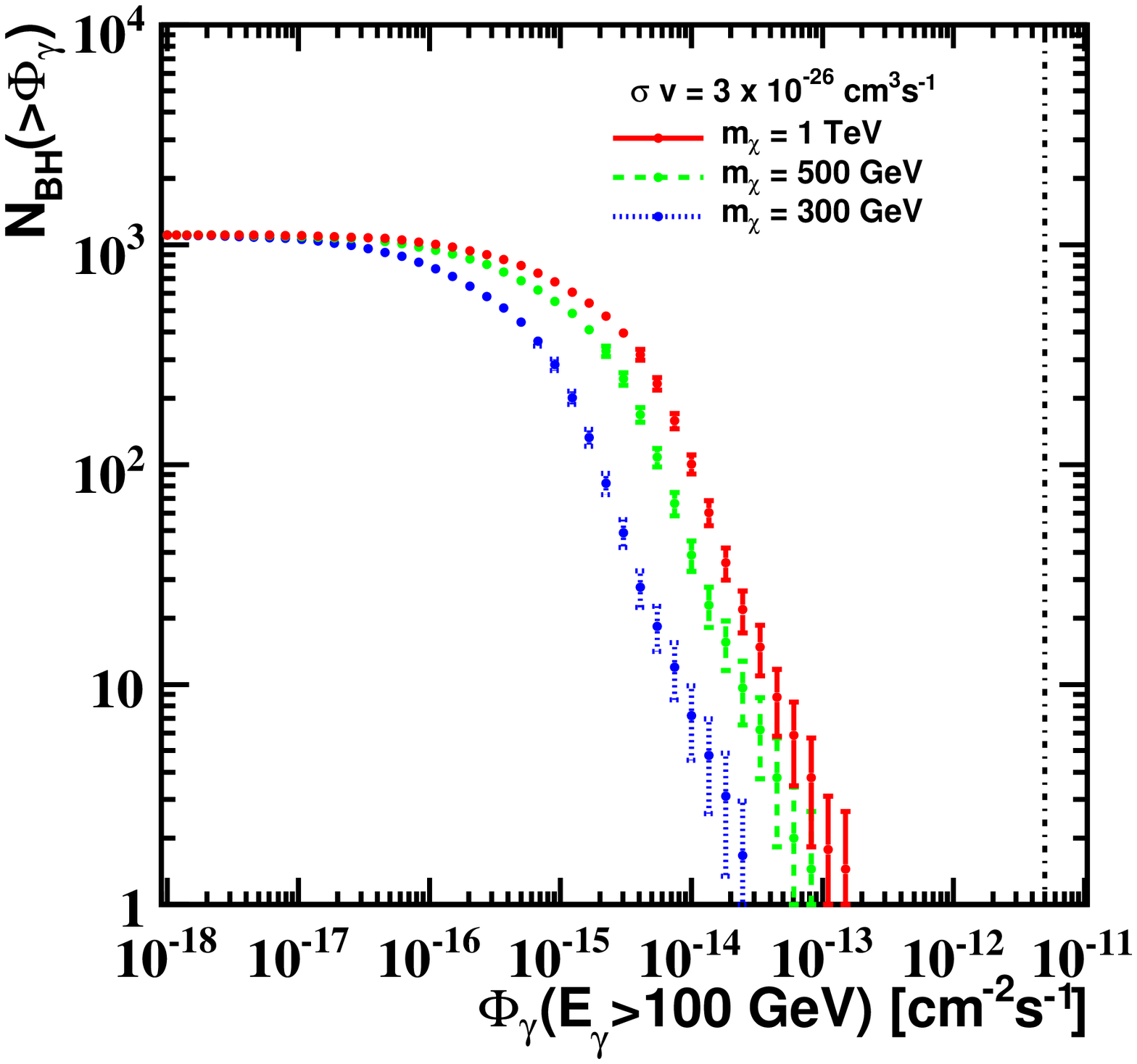}
\includegraphics[scale=0.44]{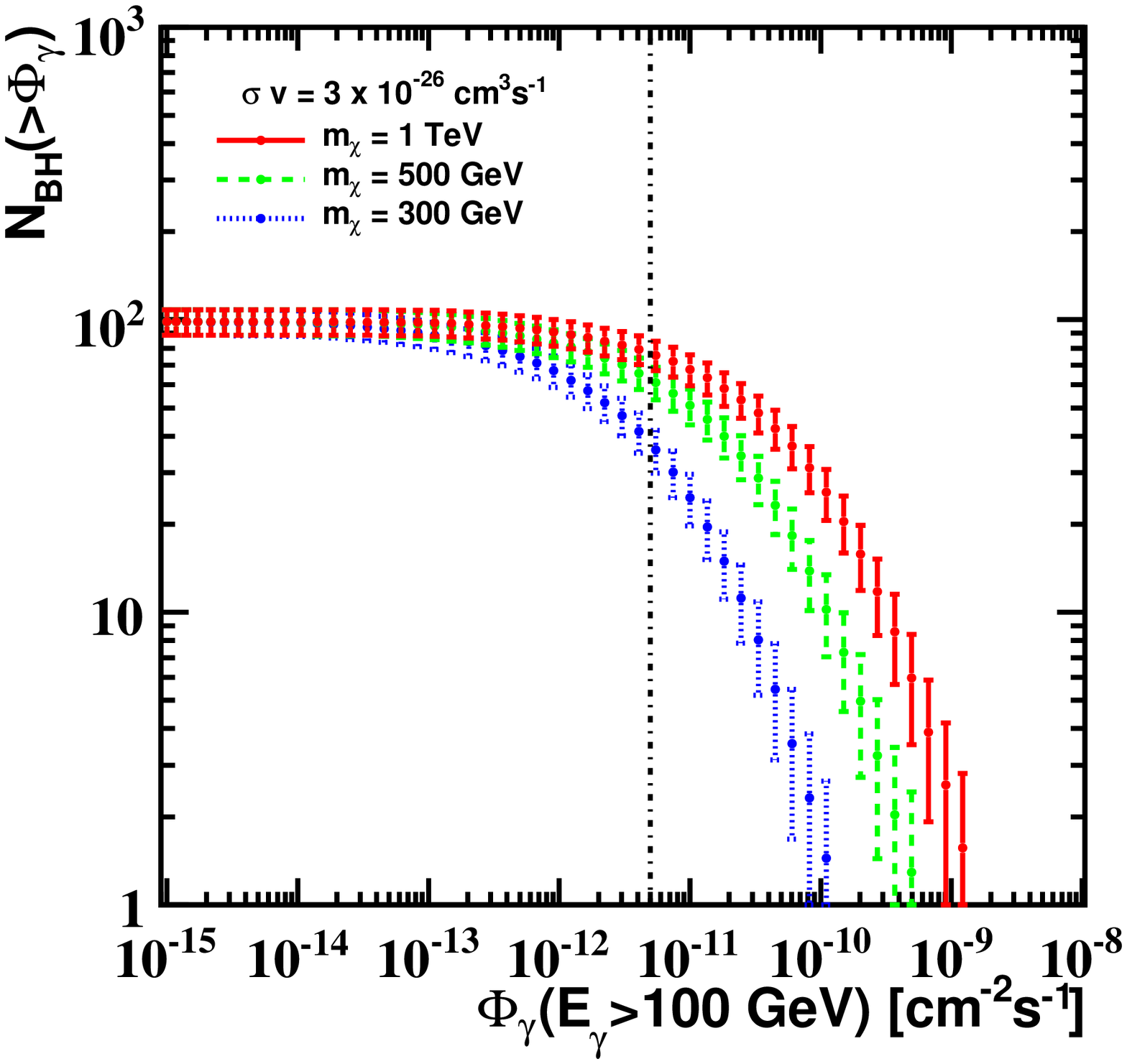}
\caption{Integrated luminosity function of IMBHs for the scenario
A, i.e. for IMBH masses of $\sim$ 10$^2$ M$_{\odot}$ (left), and
the scenario B, i.e. for IMBH masses of $\sim$ 10$^5$ M$_{\odot}$
(right), for a Milky Way-sized halo obtained from an average of
200 stochastic realizations (see text for details). Neutralino
masses of 300, 500 GeV and 1 TeV, respectively, and a
velocity-weighted annihilation cross section $\sigma v\, = \,3\,
\times\,10^{-26}\,cm^3s^{-1}$ are considered. The integrated
luminosity corresponds to the number of black holes yielding a
integrated gamma-ray flux higher than a given integrated flux,
$N_{BH}(>\Phi_{\gamma})$, as a function of the integrated flux
$\Phi_{\gamma}$. The gamma-ray flux is integrated above 100 GeV.
The nominal 5$\sigma$ H.E.S.S. point source sensitivity (25 hours)
is plotted (black dashed-dotted line) for comparison.}
\label{fig:Integrated_Luminosity}
\end{figure*}
\subsection{IMBHs formation scenarios}

IMBHs are compact objects with masses comprised between that of
the heaviest remnant of a stellar collapse, $\sim 20 M_{\odot} $
\cite{Fryer_Kalogera,Silverman}, and the lower end of the mass
range of Supermassive Black Holes (SMBH) $ \sim 10^{6} M_{\odot} $
\cite{Ferrarese_Ford,Greene}.

Observational hints of the existence of IMBHs come from the
detection of ultra-luminous X-ray sources (ULXs), apparently not
associated with active galactic
nuclei~\cite{Colbert:2002mi,Swartz:2004xt,Dewangan:2005kj}, from
stellar kinematics in globular clusters~\cite{Frank:1976,Noyola}
and emission-line time lags in galaxies~\cite{Peterson}.

From a theoretical point of view, a population of massive seed
black holes could help to explain the origin of SMBHs, especially
those powering quasars at redshift 6 or 7, that were thus already
in place when the Universe was only $\sim 1$ Gyr
old~\cite{haiman:2001}. A population of
$^{\backprime\backprime}$wandering$^{\prime\prime}$ IMBHs is in
fact a generic prediction of scenarios that seek to explain the
properties of supermassive black
holes~\cite{islama:2003,volonteri:2003,Koushiappas:2003zn}.
However, despite their theoretical interest, it is difficult to
obtain conclusive evidence for the existence of IMBHs.

In Ref.~\cite{bertone3} the consequences of the formation and
growth of IMBHs on the surrounding distribution of DM have been
studied. In particular, it was shown that these processes lead to
the formation of strong \DM\, overdensities called {\it
mini-spikes}, which are ideal targets for indirect \DM\, searches,
as they would appear as a population of gamma-ray point-sources
with identical energy spectrum.

The properties of mini-spikes have been discussed in detail for
two different scenarios. In the first one (scenario A), black
holes are remnants of the collapse of Population III (or
$^{\backprime\backprime}$first$^{\prime\prime}$)
stars~\cite{Madau:2001}, which are believed to collapse directly
to black holes in the mass range $M \sim 60 - 140 \msun$ and $M
\gtrsim 260 \msun$~\cite{Heger:2002by}. Black holes in this
scenario may not necessarily form at the very centers of their
initial host dark matter halos at high redshift, a circumstance
that, as we shall see, may have important consequences on the
detectability of IMBHs.

The second scenario (scenario B) is representative of a class of
models in which black holes originate from the collapse of
primordial gas in early-forming
halos~\cite{Haehnelt:1993,Loeb:1994,Eisenstein:1995,Haehnelt:1998,Gnedin:2001ey,Bromm:2002hb,Koushiappas:2003zn}.
The initial black holes are massive ($\sim 10^5 \msun$) and the
growth of SMBH proceeds in such a way that both mergers and
accretion play an important role.  Following Ref.~\cite{bertone3},
we focus here on the specific formation scenario proposed in
Ref.~\cite{Koushiappas:2003zn}, which goes as follows: during the
virialization and collapse of the first halos, gas cools,
collapses, and forms pressure-supported disks at the centers of
halos that are sufficiently massive to contain a relatively large
amount of molecular hydrogen. Local gravitational instabilities in
the disk lead to an effective viscosity that transfers mass inward
and angular momentum outward \cite{Lin:1987} until supernov\ae \
in the first generation of stars heat the disk and terminate this
process \cite{Koushiappas:2003zn}.  By the time the process
terminates (of the order of the lifetimes of Pop~III stars, $\sim
1-10$~Myr), a baryonic mass of order $\sim 10^5 \msun$ has lost
its angular momentum and has been transferred to the center of the
halo. Such an object may be briefly pressure-supported, but it
eventually collapses to form a black hole
\cite{Heger:2002by,Shapiro:1983}.

The characteristic mass of the black hole forming in a halo of
virial mass $M_v$ is given by \cite{Koushiappas:2003zn}
\begin{eqnarray}
\label{eq:mbh} M_{\rm bh}& = 3.8 \times 10^4 \msun \left(
\frac{\kappa}{0.5} \right)
\left( \frac{f}{0.03} \right)^{3/2} \nonumber \\
&\times \left( \frac{M_v}{10^7 \msun} \right) \left(
\frac{1+z}{18} \right)^{3/2} \left( \frac{t}{10 {\rm Myr}}
\right),
\end{eqnarray}
where  $f$ is the fraction of the total baryonic mass in the halo
that falls into the disk, $z$ is the redshift of formation,
$\kappa$ is the fraction of the baryonic mass which loses its
angular momentum and remains in the remnant black hole, and $t$ is
the timescale for the evolution of the first generation of stars
\cite{Koushiappas:2003zn}.

In order to study the consequences for indirect \DM\, searches we
will compare H.E.S.S. data with the mock catalogs of
Ref.~\cite{bertone3}, which consisted of 200 stochastic
realizations of Milky Way-like halos at $z=0$, obtained by
populating halos with black holes at high redshift (following the
prescriptions of scenarios A and B) and evolving them forward to
determine the properties of satellite black holes. The mass of the
Milky Way was fixed at $10^{12.1} h^{-1}\,M_{\odot}$ at $z=0$, and
our analysis is based on their statistically large sample of
wandering black hole populations in Milky Way-like halos of this
mass.

In scenario A, the mass spectrum of unmerged black holes is a
delta function and the average number of {\it unmerged} black
holes per Milky Way halo is $N_{\rm bh,A} \simeq 1027 \pm 84$,
where the error bar denotes the $1\sigma$ scatter from
halo-to-halo\footnote{Since merger events are likely to perturb
the DM distribution around a IMBH, only those IMBH will be
considered which never experienced a major merger.}. In scenario
B, the total number of {\it unmerged} black holes per Milky Way
halo is $N_{\rm bh,B} \simeq 101 \pm 22$~\cite{bertone3}. The
dispersion denotes the 1$\sigma$ scatter among different
realizations of Milky Way-like halos, as discussed
in~\cite{bertone3}.

\subsection{Mini-spikes}
The growth of massive black holes inevitably affects the
surrounding distribution of Dark Matter. The profile of the final
DM overdensity, called {\it mini-spike}, depends on the initial
distribution of matter, but also on astrophysical processes such
as gravitational scattering off stars and mergers.

Ignoring astrophysical effects, and assuming adiabatic growth of
the black hole (i.e. assuming that the black hole grows on a time
scale much longer than the dynamical time scales of DM around it),
one can calculate analytically the functional form of the final DM
profile. If one starts from an initially uniform DM distribution,
which is the most likely situation for black holes in scenario A,
the final profile will be a mild mini-spike with density
$\rho_{\rm sp} \propto (r/r_h)^{3/2}$ (e.g.
see~\cite{Quinlan:1995} and references therein). If one starts
from a cuspy profile that is a power-law with index $\gamma$=1, as
relevant for scenario B, the new profile is a new power-law,
\begin{equation}
\rho_{\rm sp}(r)=\rho(r_{\rm sp}) \left(\frac{r}{r_{\rm
sp}}\right)^{-\gamma_{\rm sp}}
\end{equation}
where the radius of the spike is $r_{\rm sp} \approx 0.2
\,r_h$~\cite{Merritt:2003qc}, and $\gamma_{\rm sp}$ is related to
the initial power-law index $\gamma$ by~\cite{Gondolo:1999ef}
\begin{equation}
\gamma_{\rm sp}=\frac{9-2\gamma}{4-\gamma} \;\; .
\end{equation}
In the case of the Navarro, Frenk and White profile, $\gamma=1$,
which implies $\gamma_{\rm sp}=7/3$.

To calculate the annihilation flux, the singularity of $\rho_{\rm
sp}$ at r=0 needs to be cut off; we introduce a minimal radius
r$_{\rm cut}$. One limit is given by the size of the IMBH, another
by the condition that the annihilation rate of the \DM\, particles
is smaller than the inverse age of the mini-spike :
\begin{equation}
\rho_{\rm sp}(r_{\rm  lim})=m_\chi/\sigma v \,(t-t_f) \equiv
\rho_{\rm lim} \,\,\, . \label{eq:rholim}
\end{equation}
An inner cut-off is therefore defined at a radius
\begin{equation}
r_{\rm cut}={\rm Max} \left[ 4 R_{\rm Schw}, r_{\rm lim} \right]
\end{equation}
where $R_{\rm Schw}$ is the Schwarzschild radius of the IMBH
$R_{\rm Schw} = 2.95 \,{\rm km} \, M_{\rm bh}/\msun$. For common
values of the mass and cross section of the DM particle, $r_{\rm
lim} \sim 10^{-3}$~pc so that $r_{\rm cut} = r_{\rm lim}$.

Although in principle this applies also to the black hole at the
Galactic Center, there are a number of astrophysical effects, such
as off-center formation, major mergers, and gravitational
scattering off stars, that tend to erase any DM overdensity. A
detailed discussion of the formation and evolution of the DM spike
at the Galactic center, including a discussion of the prospects
for indirect detection in light of the very high energy  (VHE,
$\rm E_{\gamma}\,>\,100\,GeV$) gamma ray source coincident with
Sgr A*, can be found in Ref.~\cite{bertone1}.

All these astrophysical processes are unlikely to take place
around IMBHs. Mini-spikes around IMBHs that {\it never experience
mergers} are therefore expected to be stable structures over
cosmological timescales, and they are thus promising targets for
indirect detection. The gamma-ray flux from these targets can be
easily calculated, once the DM profile has been determined with
the prescription outlined above. In the case of scenario B, which
leads to higher gamma-ray fluxes than scenario A, the differential
flux from a mini-spike at distance D can be written
as~\cite{bertone3}
\begin{eqnarray}
\nonumber \Phi (E,D) & = & \Phi_0 \frac{{\rm d}N}{{\rm d}E} \left(
\frac{\sigma v}{10^{-26} {\rm cm}^3{\rm s^{-1}}} \right) \left(
\frac{m_\chi}{100 {\rm GeV}} \right)^{-2}
\\
 & \times & \left( \frac{D}{{\rm kpc}} \right)^{-2}
\left( \frac{\rho(r_{\rm sp})}{10^2 {\rm GeV}{\rm cm}^{-3}}
\right)^{2}
\nonumber \\
 & \times &
\left( \frac{r_{\rm sp}}{{\rm pc}} \right)^\frac{14}{3} \left(
\frac{r_{\rm cut}}{10^{-3}{\rm pc}} \right)^{-\frac{5}{3}}\, ,
\label{eq:flux}
\end{eqnarray}
with $\Phi_0 = 9 \times 10^{-10} {\rm cm}^{-2}{\rm s}^{-1}$. The
gamma-ray spectrum per annihilation ${\rm d}N / {\rm d}E$ depends
on the nature of the DM particle, and both numerical calculations
and analytic fits are available in the literature for all possible
annihilation channels. This formula is valid for mini-spikes
forming adiabatically from an initial NFW profile, and under the
(very good) approximation $r_{\rm sp} \gg r_{\rm cut}$. In this
case, one can easily verify that  the ``luminosity'' of a
mini-spike in terms of DM annihilation is of the same order of
magnitude as that of the entire Milky Way halo.

Although one would naively
expect that the fluxes scale with $\sigma v /m_\chi^2$, 
in the
mini-spike scenario the DM profile itself depends on $m_\chi$
and $\sigma v$, and the final luminosity of the objects, for
an initially NFW profile, is
proportional to $\sim (\sigma v)^{2/7} m_\chi^{-9/7}$~\cite{bertone3}.

\begin{figure}[!ht]
\includegraphics[scale=0.42]{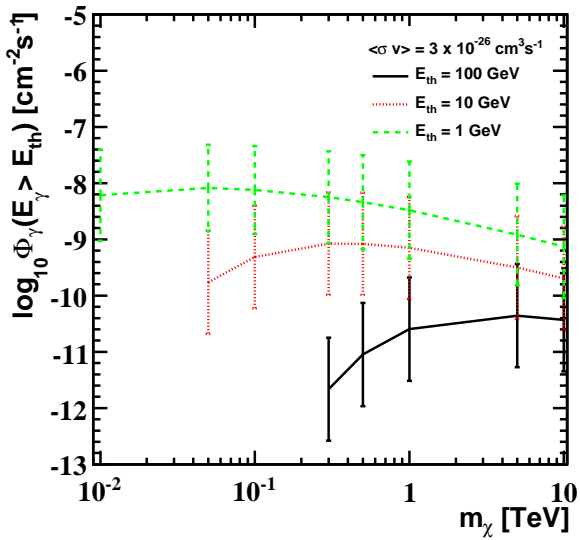}
\caption{Integrated flux $\Phi_{\gamma}$ expressed in $\rm
cm^{-2}s^{-1}$ above the energy threshold as a function of the
neutralino mass $m_{\chi}$ for thresholds of 1, 10 and 100 GeV,
respectively, and a velocity-weighted annihilation cross section
$\sigma v\, = \,3\, \times\,10^{-26}$ $\rm cm^3s^{-1}$. The quoted
error bars correspond to the r.m.s. of the integrated flux
distribution. The maximum flux is obtained for neutralino masses
well above the energy threshold of the instrument (see text for
details).} \label{fig:Integrated_Flux}
\end{figure}

Fig.~\ref{fig:Integrated_Luminosity} shows the integrated
luminosity function of IMBHs in a Milky Way-sized halo, i.e. the
number of black holes $N_{\rm BH}$ producing an integrated
gamma-ray flux higher than a given flux, as a function of the
integrated flux. This can be understood as the number of black
holes that can be detected with a telescope of given integrated
flux sensitivity. The point-source sensitivity (5$\sigma$, 25
hours at 20$^{\circ}$ zenith) for H.E.S.S. is plotted for
comparison. The integrated luminosity is shown in the case of the
aforementioned scenarios  A and B, for three different \DM\,
particle masses and an annihilation cross section $\sigma
v\,=\,3\times10^{-26}$ $\rm cm^3s^{-1}$. This value allows for the
thermal relic density of the DM particle to account for the
measured cold \DM\, density $\Omega_{\rm CDM}h^2\simeq0.1$ (see
Eq. (3.4) of Ref.~\cite{jungman}). Here, DM particles are assumed
to be neutralinos annihilating in the $b \bar{b}$ channel. For the
assumed value of $\sigma v$, a large number of IMBHs in the Milky
Way is within the reach of H.E.S.S. for scenario B. In
what follows, we will concentrate on that scenario. 
Fig.~\ref{fig:Integrated_Flux} presents the mean integrated
gamma-ray flux per IMBH for various energy thresholds as a
function of the mass of the DM particle annihilating into
$b\bar{b}$. Also displayed are the error bars corresponding to the
r.m.s. variation of the integrated flux distribution. For a 1 GeV
threshold, well suited for gamma-ray satellite experiments, the
maximum flux is obtained for a DM particle mass of $\sim$ 80 GeV.
This maximum comes from a balance between the factor
$m_{\chi}^{-9/7}$ and the integral of the annihilation spectrum up
to the DM particle mass (see Eq.~\ref{eq:flux}). Adopting an
energy threshold of 100 GeV, as appropriate for Cherenkov
telescopes such as H.E.S.S., the largest fluxes are obtained for a
mass of $\sim$ 5 TeV. For this mass, the mean of the integrated
flux distribution is $4.5\times10^{-11}$ $\rm cm^{-2}s^{-1}$. For
masses close to the experimental threshold, the integrated flux
increases with the \DM\, mass. Well above the threshold, the
standard regime is recovered, with fluxes decreasing with
$m_{\chi}$.

\section{\label{sec:hess}H.E.S.S.}
\subsection{The H.E.S.S. instrument}
The H.E.S.S. (High Energy Stereoscopic System) array is dedicated
to VHE gamma-ray astronomy. The instrument is composed of four
Imaging Atmospheric Cherenkov Telescopes located in the Khomas
Highland of Namibia at an altitude of 1800 m above sea level. This
southern location is well suited for observations towards the
inner region of the Galactic halo. Each telescope consists of an
optical reflector of 107 m$^2$ composed of 382 round
mirrors~\cite{bernlohr}. The Cherenkov light emitted by charged
particles in the electromagnetic shower initiated by the primary
gamma-ray is focused onto a camera equipped with 960
photomultiplier tubes (PMTs) of 0.16$^{\circ}$ individual field of
view~\cite{vincent}. The total field of view of H.E.S.S. is
5$^{\circ}$ in diameter. The stereoscopic technique allows for
accurate reconstruction of the direction and the energy of the
primary gamma-ray~\cite{trigger}. H.E.S.S. has an angular
resolution of about 5 arc-minutes and a source location accuracy
of $\sim$ 30'' for strong sources. The sensitivity for point-like
sources reaches $2\,\times\,10^{-13}$ $\rm cm^{-2}s^{-1}$ above 1
TeV for a 5$\sigma$ detection in 25 hours at 20$^{\circ}$
zenith~\cite{crabe}.

\subsection{Observations and data analysis}
The data used in this analysis were collected between 2004 and
2007 during the Galactic plane survey with the four telescope
H.E.S.S. array. The Galactic plane has been observed between $\pm$
3$^{\circ}$ in latitude and from -30$^{\circ}$ to 60$^{\circ}$ in
longitude relative to the Galactic Center  in 884 pointings.  Runs
are taken mainly with 28 minute duration at pointing positions
with a typical spacing of 0.4$^{\circ}$ in longitude and
1$^{\circ}$ in latitude. For the analysis described in this paper,
the observations dedicated to known sources at other wavelengths
are excluded. Astrophysical models of TeV emission are available
for these sources so that the search is focused on regions where
no standard astrophysical emitters have been detected by H.E.S.S.
After the standard quality selection procedure~\cite{crabe} and
dead time correction, the data set amounts to $\sim$ 400 hours of
live time and a mean zenith angle of all observations is $\sim$
30$^{\circ}$ resulting in a typical energy threshold of 200 GeV.

Following the standard calibration of the shower images from PMT
signals~\cite{aharonian2}, the event reconstruction scheme using
the combined \textit{Model-Hillas} analysis is applied to the data
to select the gamma events. The \textit{Hillas} reconstruction is
based on the Hillas geometrical moment of the
image~\cite{aharonian3}. The \textit{Model} analysis is based on
a pixel-by-pixel comparison of the image to a template image
generated by a semi-analytical model of the
shower~\cite{denauroi0}. Both methods yield a typical energy
resolution of 15\% and an angular resolution at 68\% full
containment radius better than 0.1$^{\circ}$. The data analysis is
done with a combination of these two methods to improve the
hadronic background rejection~\cite{aharonian3}. An additional cut
on the primary interaction depth is also used to improve
background rejection. After a cleaning of the images, the
direction, energy, impact parameter and primary interaction point
are reconstructed for each gamma event. A cut on the image size at
60 photoelectrons is used to obtain better sensitivity to weak
gamma-ray excesses. The background level is estimated using the
template model method~\cite{rowell} as described below. This
allows to estimate the background level at each sky position.
\begin{figure*}[!ht]
\mbox{\hspace{-0.5cm}\includegraphics[scale=0.95]{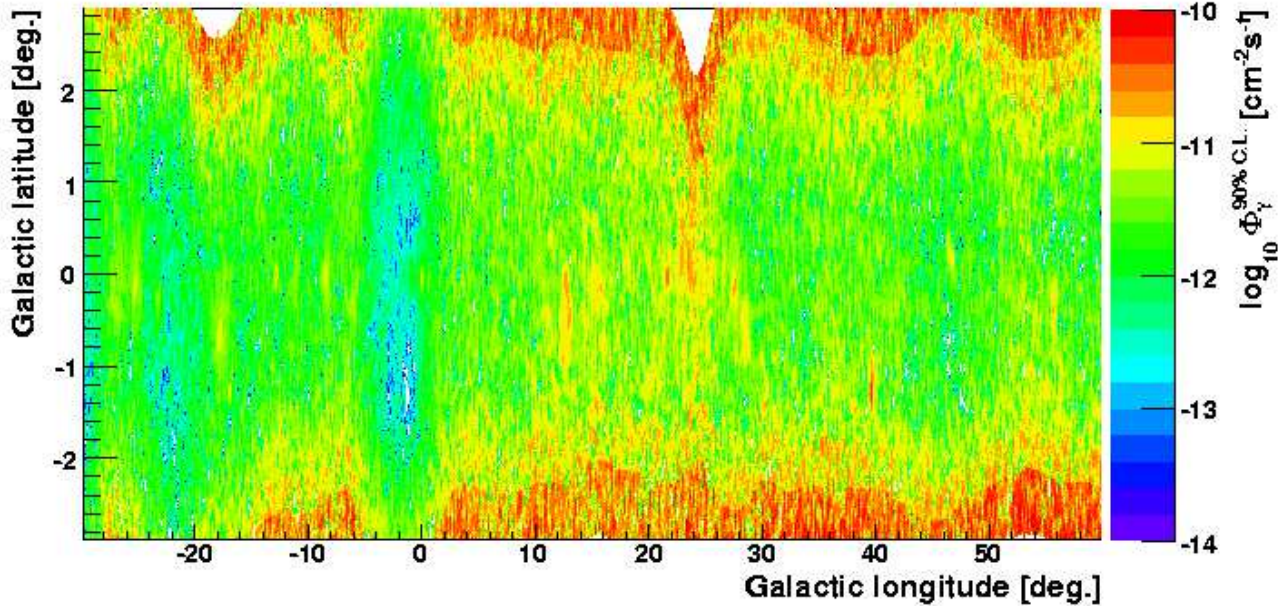}}
\caption{H.E.S.S. sensitivity map in Galactic coordinates, i.e.
90\% C.L. limit on the integrated gamma-ray flux above 100 GeV,
for \DM\, annihilation assuming a \DM\, particle of mass
m$_{\chi}$ = 500 GeV and annihilation into the $b\bar{b}$ channel.
The flux sensitivity is correlated to the exposure and acceptance
maps. In the Galactic latitude band between -2$^{\circ}$ and
2$^{\circ}$, the gamma-ray flux sensitivity reaches $10^{-12}$
$\rm cm^{-2}s^{-1}$.} \label{fig:HESS_Sensitivity_Map}
\end{figure*}

\subsection{Gamma-ray sensitivity map}
In order to study the H.E.S.S. sensitivity in the Galactic plane
survey  to \DM\, annihilations from mini-spikes, we perform the
analysis on the Galactic plane survey data. The H.E.S.S. flux
sensitivity map is derived using the following sky maps in
Galactic coordinates which are estimated within the survey range:
\begin{itemize}
\item the observed gamma-ray map,  \item the normalized measured
background map, \item the effective time exposure map (in
seconds), \item the gamma-ray acceptance map (in m$^2$).
\end{itemize}
\noindent The gamma-ray map is obtained after the event selection
and reconstruction procedures described above. The sky map is
divided into bins of 0.02$^{\circ}$ $\times$ 0.02$^{\circ}$ size
and an oversampling radius of 0.1$^{\circ}$ is applied to the
data. The oversampling smoothes the map with a top-hat function of
radius of 0.1$^{\circ}$ to match the point spread function of the
instrument. In each bin of the map in Galactic coordinates, the
background rate is estimated by the template background method
(see~\cite{rowell} for details) and the normalized background map
is obtained by the following relation:
\begin{equation}
N_{\rm Bkg}^{\rm Norm}(b,l)\,=\,N_{\rm Bkg}(b,l)\times
\frac{Acc_{\gamma}(b,l)}{Acc_{\rm h}(b,l)}
\end{equation}
where the acceptances for gamma-ray-like events, $Acc_{\gamma}$,
and hadron-like events, $Acc_{\rm h}$, are computed in each
position of Galactic longitude
$b$ and latitude $l$ taking into account 
collection area corrections. Using the observed gamma-ray map and
the normalized background map, we derive a sensitivity map
expressed in terms of an upper limit (at 90\% C.L) on the number
of gamma events above the nominal energy threshold in each sky
position. Using the exposure and acceptance maps, the flux
sensitivity map for gamma-rays from \DM\, annihilation is then
determined by :
\begin{eqnarray}
\label{eqn:phi} \lefteqn{\Phi_{\gamma}^{90\% C.L}(b,l)\,=\,}
\nonumber
& & \\
& &
\hspace{-1.0cm}\frac{N_{\gamma}^{90\%C.L}(b,l)\,\displaystyle\int_{E_{\rm
th}}^{m_{\rm
DM}}\frac{dN}{dE_{\gamma}}\Big(E_{\gamma}\Big)\,dE_{\gamma}}{\displaystyle\int_{T_{\rm
obs}}\int_{0}^{m_{\rm DM}} A_{\rm
eff}\Big(E_{\gamma},z(b,l),\theta(b,l)\Big)\frac{dN}{dE_{\gamma}}\Big(E_{\gamma}\Big)\,dE_{\gamma}\,d\tau}.
\end{eqnarray}
Here, A$_{\rm eff}$ is the effective area for gamma-rays which is
a function of the gamma energy E$_{\gamma}$ and the sky position
$(b,l)$. The d$\tau$ integration denotes the averaging according
to the zenith angle $z$ and offset $\theta$ distribution of the
observation livetime. The annihilation spectrum of the \DM\,
particle of mass $m_{\rm DM}$ is denoted by $dN/dE_{\gamma}$ and
is integrated from the nominal energy threshold $E_{\rm th}$,
which is about 100 GeV in the survey, up to $m_{\rm DM}$. The
integral over $\tau$ is calculated over the total observation
livetime $T_{\rm obs}$.

Fig.~\ref{fig:HESS_Sensitivity_Map} shows the experimentally
observed sensitivity map in the Galactic plane from Galactic
longitudes $l=-30^{\circ}$ to $l=+60^{\circ}$ and Galactic
latitudes $b=-3^{\circ}$ to $b=+3^{\circ}$, for a \DM\, particle
of 500 GeV mass annihilating into the $b\bar{b}$ channel. The
H.E.S.S. sensitivity depends strongly on the exposure time and
acceptance maps which are related to the choice of the pointing
positions. The flux sensitivity varies along the latitude and
longitude due to inhomogeneous coverage of the Galactic plane. In
principle, the sensitivity map depends on the \DM\, annihilation
spectrum. However, as can be seen from Eq.~(\ref{eqn:phi}), the
spectrum is balanced by the effective area which mainly drives the
result of the integral. The particle mass is not expected to bring
about strong variations in the map as long as the mass is larger
than 100 GeV.

In the band between $-2^{\circ}$ and $2^{\circ}$ in Galactic
latitude, a DM annihilation flux sensitivity at the level of
$10^{-12}$ $\rm cm^{-2}s^{-1}$ is achieved. H.E.S.S. thus reaches
the required sensitivity to be able to test \DM\, annihilations
from mini-spikes in the context of one relatively favorable
scenario for IMBH formation and adiabatic growth of the DM halo
around the black hole.

Deeper observations of the GC and at Galactic longitude of $\sim
-20^{\circ}$ allow the flux sensitivity of $\sim 5\times10^{-13}$
$\rm cm^{-2}s^{-1}$ for a 500 GeV DM particle annihilating in the
$b\bar{b}$ channel. For $b=0^{\circ}$ and $l=-6^{\circ}$, the flux
sensitivity is $\sim 10^{-12}$ $\rm cm^{-2}s^{-1}$. For $|b|\ge
2^{\circ}$, the sensitivity is deteriorated due to a weaker
effective exposure. For $b=0^{\circ}$ and $l=-0.5^{\circ}$, near
the Galactic Center, the flux sensitivity is $\sim 10^{-13}$ $\rm
cm^{-2}s^{-1}$ in the 100\% BR $b\bar{b}$ annihilation channel and
$\sim 5\times 10^{-14}$ $\rm cm^{-2}s^{-1}$ in the $\tau^+\tau^-$
channel.

\section{Results and Discussion}
H.E.S.S. observations (2004-2006) of the Galactic plane allowed to
discover more than 20 VHE sources~\cite{sp2}. Some of them have
been identified owing to their counterparts at other wavelengths,
but almost half of the sources have no obvious counterpart and are
still unidentified~\cite{darksources}. For these sources, an
accurate reconstruction of their energy spectra has been carried
out. All spectra were consistent with a pure power-law of spectral
indices between 2.0 and 2.5, spanning up to two orders of
magnitude in energy above the energy threshold, as shown in Fig.~8
of Ref.~\cite{darksources}. None of them exhibits an energy
cut-off, characteristic of \DM\, annihilation spectra, in the
energy range from $\sim$ 100 GeV up to 10 TeV. Furthermore, the
detailed study of their morphology~\cite{darksources} show that
all the sources have an intrinsic spatial extension greater than
$\sim$5', while mini-spikes are expected to be point-like sources,
since the bulk of the gamma-ray emission comes from a region of
size $\approx 10^{-3}$ pc  at a typical distance from GC of 10
kpc. In the survey region discussed here, only three point-like
gamma-ray sources are detected: HESS J1826-148, HESS J1747-281 and
HESS J1745-290. They are identified with well-known objects: LS
5039, G0.9+0.1 and the Galactic Center, respectively. H.E.S.S. has
detected so far no IMBH candidate within the survey range.

\begin{figure}[!ht]
\mbox{\hspace{-0.cm}\includegraphics[scale=0.45]{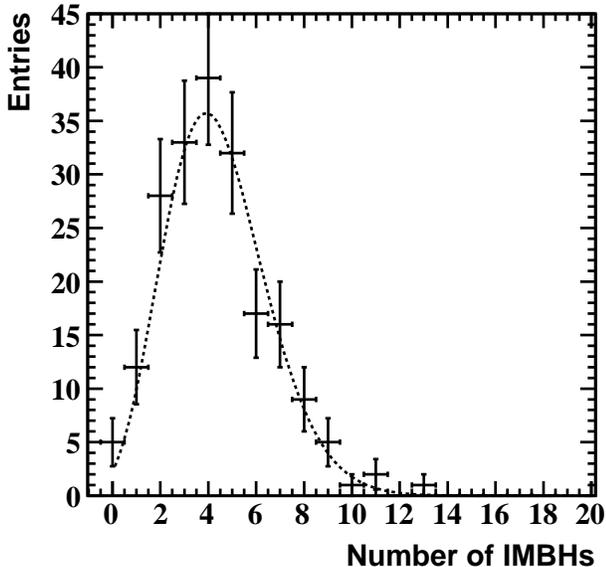}}
\caption{Distribution of the number of IMBHs for the scenario B in
the survey range corresponding to a field of view between
$\pm$3$^{\circ}$ in Galactic latitude and from -30$^{\circ}$ to
60$^{\circ}$ in Galactic longitude. The distribution is obtained
from 200 stochastic realizations of Milky Way-like halos (see Sec.
II. A for details). The mean of the distribution is 4.3 with an
rms of 2.3. The distribution is well fitted to a Poisson
distribution with mean 4.4 (black dashed line).}
\label{fig:Distribution_IMBH_SurveyRange}
\end{figure}

\begin{figure}[!hb]
\mbox{\hspace{-0.cm}\includegraphics[scale=0.45]{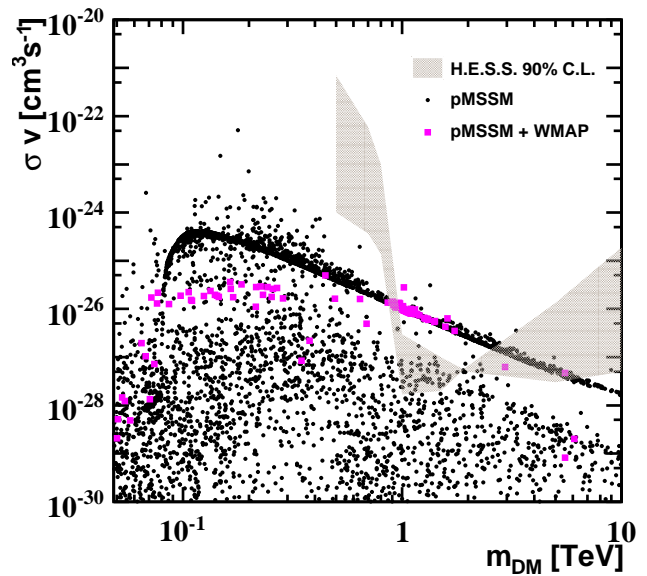}}
\caption{Constraints on the IMBH gamma-ray production scenario for
different neutralino parameters, shown as upper limits on the
annihilation cross section $\sigma v$ as a function of the mass of
the \DM\, particle $m_{\rm DM}$, but with a number of implicit
assumptions about the IMBH initial mass function and halo profile
(see text for details). For scenario B, the probability of having
no observable halos in our Galaxy is 10\% from Poisson statistics
making these limits essentially 90\% confidence level exclusion
limits for this one particular (albeit optimistic) scenario (grey
shaded area). The \DM\, particle is assumed to be a neutralino
annihilating into $b\bar{b}$ pairs or $\tau^+\tau^-$ pairs to
encompass the softest and hardest annihilation spectra. The limit
is derived from the H.E.S.S. flux sensitivity in the Galactic
plane survey within the mini-spike scenario. SUSY models from the
pMSSM (black points) are plotted together with those satisfying
the WMAP constraints on the \DM\, particle relic density (magenta
points).} \label{fig:exclusionlimit_pMSSM}
\end{figure}

\begin{figure}[ht]
\mbox{\hspace{-0.cm}\includegraphics[scale=0.45]{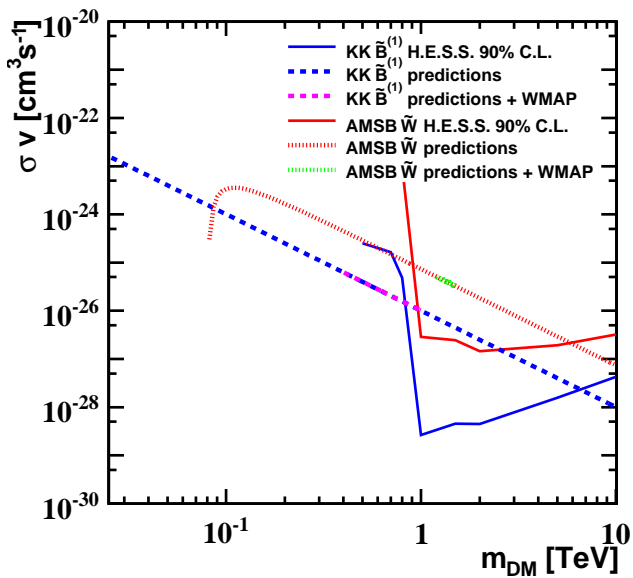}}
\caption{90\% C.L. exclusion limits on the annihilation cross
section $\sigma v$ as a function of the mass of the \DM\, particle
$m_{\rm DM}$ for a Kaluza-Klein boson $\tilde{B}^{(1)}$ (solid
blue line) and a pure wino $\tilde{W}$ in AMSB (solid red line).
The limit is derived from the H.E.S.S. flux sensitivity in the
Galactic plane survey within the mini-spike scenario. 
Kaluza-Klein models in UED scenario (blue dashed line) and AMSB
models (red dotted line) are plotted together with those
satisfying the WMAP constraints on the \DM\, particle relic
density (pink dashed line and green dotted line, respectively).}
\label{fig:exclusionlimit_KK_AMSB}
\end{figure}

Based on the absence of plausible IMBH candidates in the H.E.S.S.
data, we can derive constraints on one scenario for neutralino or
LKP dark matter annihilations. These constraints are shown as
upper limits on the annihilation cross section
(Figures~\ref{fig:exclusionlimit_pMSSM}
and~\ref{fig:exclusionlimit_KK_AMSB}) but are actually constraints
on the entire gamma-ray production scenario, assuming the mass
function of scenario B and a coreless halo model. In the survey
range, the expected mean number of IMBHs is 4.3, within the
assumptions of scenario B, as shown in
Fig.~\ref{fig:Distribution_IMBH_SurveyRange}. The number of IMBHs
in the whole halo depends on the assumed cosmological parameters
in the formation scenario~\cite{Koushiappas:2005qz}, in particular
the reionization redshift. The value of 4.3 {\it unmerged} IMBHs
is for a reionization redshift of 16. The redshift of $10.8\pm1.4$
determined from latest WMAP data~\cite{Komatsu:2008hk} favors a
larger number of IMBHs since they had less time for merger events
to occur~\cite{Koushiappas:2005qz}.
The distribution of the IMBH number has an intrinsic rms of 2.3.
The probability to find no IMBH in the survey range is 2.5\%. The
simulations of the Milky Way halo show that the IMBH number
distribution in the sky survey range is well fitted with a Poisson
distribution with mean 4.4.

For a given gamma-ray flux sensitivity $\Phi$, we define the
probability density function of black holes yielding an integrated
gamma-ray flux larger than $\Phi$, $d^2N_{BH}(>\Phi)/dbdl$, from
the IMBH catalog used in~\cite{bertone3} by :
\begin{eqnarray}
\label{eqn:nbhdetected} \lefteqn{\,\,\,\,\,\,\,N_{\rm BH}({\rm
detected},>\Phi)\,=\,} \nonumber
& & \\
& &
\int_{b=-3^{\circ}}^{b=+3^{\circ}}\int_{l=-30^{\circ}}^{l=+60^{\circ}}
\frac{d^2N_{\rm BH}(>\Phi)}{dbdl}\,dbdl\,.
\end{eqnarray}
$N_{BH}({\rm detected},>\Phi)$ denotes the expected number of
black holes yielding a gamma-ray flux larger than $\Phi$ in the
survey, $\Phi$ corresponding to the H.E.S.S. sensitivity. The
total number of black holes in the survey is computed by
integrating the probability density function $d^2N_{\rm
BH}(>\Phi)/dbdl$ over the latitude and longitude ranges of the
Galactic plane survey. Since no H.E.S.S. source is a plausible
IMBH candidate, we thus calculate for each \DM\, particle mass the
limit at 90\% C.L. on the velocity-weighted cross section $\sigma
v$, assuming a Poisson distribution.

Fig.~\ref{fig:exclusionlimit_pMSSM} shows the exclusion limit at
the 90\% C.L. on $\sigma v$ as a function of the neutralino mass
$m_{\rm DM}$. The neutralino is assumed to annihilate into
$b\bar{b}$ and $\tau^+\tau^-$ with 100\% BR, respectively.
Below $\sim$ 2 TeV, the upper contour of the gray shaded area is
given by the $b\bar{b}$ assumption and yields the conservative
exclusion limit. Above 2 TeV, the upper contour corresponds to the
$\tau^+\tau^-$ annihilation spectrum. The maximum sensitivity of
H.E.S.S. to DM annihilation for the $b\bar{b}$ channel is achieved
for masses of order 5 TeV as expected from
Fig.~\ref{fig:Integrated_Flux}. For neutralino masses in the TeV
energy range, we obtain limits on one mini-spike scenario
(scenario B). The limits on $\sigma v$ are at the level of
$10^{-28}$ $\rm cm^{-3}s^{-1}$ for the $b\bar{b}$ channel. A rapid
decrease in sensitivity is observed for WIMP masses less than
$\sim$~800 GeV. This corresponds to the threshold effect seen in
Fig.~\ref{fig:Integrated_Flux}.
\begin{table}[!ht]
\caption{\label{tab:parameterspace} pMSSM parameter space randomly
scanned to generate SUSY models. A set of parameters corresponds
to a specific pMSSM model.}
\begin{ruledtabular}
\begin{tabular}{ccc}
Parameter&Minimum& Maximum\\
\hline
\\
$m_0$ &1 TeV& 30 TeV\\
$M_2$ &1 TeV &50 TeV\\
$\mu$ &1 TeV& 50 TeV\\
$m_A$ &1 TeV&50 TeV\\
$A_{t,b}$ &-300 GeV&+300 GeV\\
$tan \beta$ &3& 60\\
\end{tabular}
\end{ruledtabular}
\end{table}
Predictions for SUSY models are generated using the DarkSUSY
code~\cite{darksusy} in a phenomenological MSSM (pMSSM) framework
characterized by the following independent parameters: the common
scalar mass $m_0$, the gaugino mass parameter $M_2$, the higgsino
mass parameter $\mu$, the CP-odd Higgs mass $M_A$, the trilinear
couplings A$_{t,b}$, and the Higgs vacuum expectation value ratio
$tan \beta$. The set of parameters for a given model is randomly
chosen in the pMSSM parameter space encompassing a large class of
pMSSM models, as described in Table~\ref{tab:parameterspace}.
Models providing a neutralino thermal relic density, $\Omega_{\rm
DM}h^2$, in the range [0.08,0.12] are overlaid to account for the
cold dark matter density inferred from the measurements of the CMB
anisotropies of the Wilkinson Microwave Anisotropy Probe (WMAP)
satellite. Some models in the mass range 0.8 - 6 TeV can be
excluded.

Fig.~\ref{fig:exclusionlimit_KK_AMSB} shows the 90\% C.L.
exclusion limit on $\sigma v$ as a function of the DM particle
mass $m_{\rm DM}$ in the case of AMSB and Kaluza-Klein scenarios.
In both cases, the BRs of each annihilation channel entering in
the calculation of the total annihilation differential spectra are
uniquely determined as discussed in section II. In the AMSB
scenario, the neutralino is considered here to be a pure wino
annihilating with 100\% BR into the $W^+W^-$ channel. The
predictions for $\sigma v$ are parameterized using the results
of~\cite{moroi} where it is derived in the case of the latter
assumption on the annihilation scheme. The AMSB models fulfilling
in addition the WMAP constraints are overlaid. Models yielding
neutralino masses between 900 GeV and 6 TeV are excluded including
those having a neutralino thermal relic density compatible with
the WMAP measurements. Predictions on the annihilation cross
section for pairs of $\tilde{B}^{(1)}$ from UED
theories~\cite{sigmavKK} are also plotted as well as those
satisfying the WMAP constraints. In this scenario,
$\tilde{B}^{(1)}$ masses in the range from 0.8 to 6 TeV can be
excluded.

\section{Summary}
Observational clues for the existence of IMBHs start to
accumulate. If they indeed exist, IMBH could be \DM\, annihilation
boosters. The prospects for detecting \DM\, annihilation around
IMBHs have been widely discussed in the
literature~\cite{Brun:2007tn,Fornasa:2007ap,Horiuchi:2006de,bertone3,bertone4,zhao}.
In this work, we derive the first experimental constraints on the
mini-spike scenario of Ref~\cite{bertone3}.

Using H.E.S.S. data collected in the Galactic plane survey, we
show that H.E.S.S. has the required sensitivity to probe
gamma-rays from \DM\, annihilation in mini-spikes around IMBHs
believed to populate the Milky Way halo. The new analysis using
$\sim$ 400 hours of data taken in the Galactic plane but not
foreseen initially for this purpose, allows to derive flux
sensitivity limits for indirect \DM\, search. Combining all the
survey data, the gamma-ray flux sensitivity map is derived for
\DM\, annihilation in the region $\rm [-30^{\circ},60^{\circ}]$ in
longitude and $\rm [-3^{\circ},3^{\circ}]$ in latitude. We show
that H.E.S.S. reaches a flux sensitivity of $\sim10^{-12}$ $\rm
cm^{-2}s^{-1}$ above 100 GeV.

For the first time, a clumpiness scenario has been tested in a
large field of view with an IACT. Strong constraints are obtained
in one of the two IMBH formation scenarios
discussed~\cite{bertone3} (scenario B). The absence of plausible
candidates for Galactic IMBHs in the H.E.S.S. Galactic plane data
set allows us to put constraints on one of the more optimistic
scenarios for detecting neutralino or LKP annihilation from
mini-spikes around IMBHs. The first experimental exclusion limits
at 90\% C.L. on the velocity-weighted annihilation cross section
as a function of the \DM\, particle mass within the mini-spike
scenario are derived. Predictions from various WIMP particle
physics scenarios are constrained.

Since the characteristic annihilation flux from a IMBH varies as
$(\sigma v)^{2/7}$, reflecting the depletion of the inner part of
dark matter halo due to annihilation during the lifetime of the
IMBH, limits derived on $\sigma v$ in the absence of a detection
are proportional to $(\Phi_{\rm min}/\Phi_{\rm ref})^{7/2}$, where
$\Phi_{\rm min}$ is the sensitivity limit and $\Phi_{\rm ref}$ the
predicted annihilation flux for a nominal value of $\sigma v$.
Uncertainties in $\Phi_{\rm ref}$, reflecting for example the
imperfect knowledge of the exact shape of the dark matter spike
close to the IMBH, cause correspondingly enlarged uncertainties on
the limits on $\sigma v$. Similarly, variations in the predicted
number and mass of IMBHs may influence the limits; for example, at
least about 50 IMBHs have to be contained in the Galactic halo in
order to set a meaning flux upper limit, given the limited solid
angle coverage of the H.E.S.S. survey. The limits given here apply
within the formalism and approximations of scenario B
of~\cite{bertone3}.

The analysis described here can be adapted to other particle
physics parameters, e.g. different types of DM particles and
annihilation channels. Constraints are derived within the
mini-spike scenario although the method developed here is generic
and is suited whatever the assumed dark matter clump scheme. The
flux
sensitivity map 
is a powerful tool to investigate the sensitivity to other types
of overdensities such as small scale
clumps~\cite{diemand,koushiappas,pieri,bergstrom1} or compact
\DM\, structures like
spikes~\cite{Gondolo:1999ef,bertone1,bertone5,ullio,bertone6}.\\

\begin{acknowledgments}
The support of the Namibian authorities and of the University of
Namibia in facilitating the construction and operation of H.E.S.S.
is gratefully acknowledged, as is the support by the German
Ministry for Education and Research (BMBF), the Max Planck
Society, the French Ministry for Research, the CNRS-IN2P3 and the
Astroparticle Interdisciplinary Programme of the CNRS, the U.K.
Particle Physics and Astronomy Research Council (PPARC), the IPNP
of the Charles University, the South African Department of Science
and Technology and National Research Foundation, and by the
University of Namibia. We appreciate the excellent work of the
technical support staff in Berlin, Durham, Hamburg, Heildelberg,
Palaiseau, Paris, Saclay, and in Namibia in the construction and
operation of the equipment. We thank Andrew Zentner for making the
IMBH catalog available for this work.
\end{acknowledgments}

\end{document}